\documentclass[aps,prb,twocolumn,showpacs,groupedaddress]{revtex4}
\def\bc{\begin{center}}
\def\ec{\end{center}}
\def\be{\begin{equation}}
\def\ee{\end{equation}}
\def\bea{\begin{eqnarray}}
\def\eea{\end{eqnarray}}
\renewcommand{\vec}[1]{\mbox{\boldmath$#1$}}
\usepackage{longtable}
\usepackage{epsfig}

\begin{document}

\title{Berry phases for composite fermions: effective magnetic field and fractional 
statistics}
\author{Gun Sang Jeon, Kenneth L. Graham, and Jainendra K. Jain}
\affiliation{Physics Department, 104 Davey Laboratory, The Pennsylvania State University,
University Park, Pennsylvania 16802}

\date{\today}

\begin{abstract}
The quantum Hall superfluid is presently the only viable candidate for 
a realization of quasiparticles with fractional Berry phase statistics.
For a simple vortex excitation, relevant for a subset of fractional 
Hall states considered by Laughlin, non-trivial Berry phase statistics were 
demonstrated many years ago by Arovas, Schrieffer, and Wilczek.
The quasiparticles are in general more complicated, described accurately 
in terms of excited composite fermions.  We use the method developed by 
Kj{\o}nsberg, Myrheim and Leinaas to compute the Berry phase 
for a single composite-fermion quasiparticle, and find that it 
agrees with the effective magnetic field concept for composite fermions.  
We then evaluate the ``fractional statistics," related to the change  
in the Berry phase for a closed loop caused by the insertion of 
another composite-fermion quasiparticle in the interior.
Our results support the general validity of fractional statistics in the 
quantum Hall superfluid, while also 
giving a quantitative account of corrections to it when the quasiparticle 
wave functions overlap.  Many caveats, both practical and conceptual, are mentioned 
that will be relevant to an experimental measurement of 
the fractional statistics.  A short report on some parts of this article 
has appeared previously.
\end{abstract}
\maketitle

\section{introduction}

When hard-core particles
are confined in two dimensions, the configuration space is multiply
connected, and paths with different winding numbers are
topologically distinct because they cannot be continuously deformed into one another.
The particles are said to have statistics $\theta^*$ if a path 
independent phase $2\pi\theta^*$ results 
when one particle traverses around another in a complete loop.
A half loop is equivalent to an exchange of particles, assuming 
translational invariance, which produces a phase factor 
$e^{i\pi \theta^*}=(-1)^{\theta^*}$.  As pointed out by Leinaas and 
Myrheim,~\cite{Leinaas} non-integral values of $\theta^*$ 
are allowed, which are referred to as fractional statistics.
Clearly, fractional statistics can  
be consistently defined only in two dimensions, because in higher dimensions the  
notion of a particle going around another is topologically ill-defined.

Given the experimental fact that all fundamental particles in nature 
are either bosons or fermions, any discussion of fractional statistics 
may appear academic.  That would perhaps be true from an elementary particle 
physicist's perspective.  However, there is no fundamental principle that precludes
the possibility that certain {\em emergent} quasiparticles of a condensed matter system
might possess fractional statistics; indeed, an appearance of such statistics 
would be an interesting example of how entirely new concepts can 
emerge~\cite{Anderson} in a many body system.
Obviously, it would take a highly non-trivial state of matter in order for 
fractional statistics particles to emerge, and nature has graciously 
provided a possible candidate, namely the quantum Hall superfluid
(QHS).~\cite{Tsui}
The QHS is formed when interacting electrons confined 
to two dimensions are exposed to a strong magnetic field.  It is characterized
by quantized Hall plateaus and a vanishing resistance at  
zero temperature (in spite of the presence of disorder).
The investigation of the QHS has given rise to much interesting 
physics since its discovery in the early 1980's.~\cite{Tsui}

The possibility of fractional statistics in the QHS  
was first recognized by Halperin,~\cite{Halperin}
demonstrated in a microscopic theory by Arovas, Schrieffer, 
and Wilczek~\cite{Arovas} for a ``vortex" excitation in Laughlin's wave 
function~\cite{Laughlin} at filling factors of the form $\nu=1/m$, 
$m$ odd,  and argued to be a general 
consequence of incompressibility at a fractional filling of the lowest Landau 
level.~\cite{Su}

It is by now clear that the physics of the QHS can be 
understood, both qualitatively and quantitatively, 
without any mention of fractional statistics.  Jain showed~\cite{Jain}
that the non-perturbative physics of the QHS lies in the formation of  
particles that are {\em fermions}, called composite fermions, which are bound 
states of electrons and an even number of quantized vortices.  Many  
essential properties of the QHS can be 
explained by neglecting the interaction between composite fermions
altogether, as properties of almost free fermions. 
Extensive experimental and theoretical work has 
established that the composite-fermion (CF) theory 
gives a {\em complete} description of the low-energy Hilbert space of 
the system,~\cite{Heinonen,JK,Dev,Dev2} in that it correctly predicts the 
quantum numbers of the low-energy states and also gives accurate microscopic 
wave function for them.  
Thus, neither the explanation of the phenomenology nor a calculation of 
the experimentally measurable quantities 
requires, in principle, any consideration of fractional statistics.

With the exception of the ``quasihole" at $\nu=1/m$, 
the excitations of fractional-quantum-Hall-effect (FQHE) are not 
described by a simple vortex, but have 
a much more complicated structure, just as the FQHE 
ground states in general have much more complex wave functions than 
those at $\nu=1/m$.  Analytical calculations for the Berry phase 
statistics have not been possible for the non-vortex excitations, but 
numerical calculations have been carried out and showed surprising results.
For the {\em quasiparticles} (as opposed to quasiholes)
at $\nu=1/m$, a calculation by Kj{\o}nsberg and Myrheim,~\cite{Kjonsberg1} with 
a trial wave function suggested by Laughlin~\cite{Laughlin} showed that they do  
{\em not} possess a well defined fractional statistics,
in the sense that the calculated statistics parameter shows rapid variations 
with the separation between the two quasiparticles and fails to reach 
an asymptotic limit.  That appears to cast doubt on the generality of the 
concept of fractional statistics, and also on the validity of the earlier 
heuristic derivations that were based on 
nothing more than incompressibility at a fractional filling.~\cite{Su}

The microscopic understanding in terms 
of composite fermions has enabled further progress.
Because the CF theory gives a good account of the 
low-energy physics, it must also contain information about fractional
Berry phase statistics, provided it really exists. 
One might naively think that the fractional statistics is 
incompatible with the CF theory, but that is not the case.
As discussed by Goldhaber and Jain,~\cite{Goldhaber} the CF theory, in fact,  
provides a straightforward heuristic ``derivation" for 
fractional statistics.
The fractional statistics simply keeps track of how the effective 
magnetic field experienced by composite fermions is affected by 
local deformations in the density, as obtained, for example, 
by creation of a localized excitation. 
The CF theory allows one to go beyond the simple 
vortex at $\nu=1/m$ through the general understanding of quasiparticles 
as excited composite fermions. 
The wave function for the CF quasiparticle at $\nu=1/m$
written by Jain~\cite{Jain} 
is known to be more accurate than the one suggested by 
Laughlin.~\cite{Kasner,Girlich,Bonesteel,Quasiparticles} 
An important step in the clarification of the issue 
of fractional statistics was taken by Kj{\o}nsberg and Leinaas, who showed 
that when the former wave function is used for a calculation of the 
statistics, a definite value is obtained.~\cite{Kjonsberg2}  
The present study confirms that the result is robust to 
projection into the lowest Landau level, sorts out 
certain subtle corrections left out in the earlier study, and 
extends the calculation to more complex excitations of other 
incompressible states. 
A brief account of some of the results below has 
appeared previously in a short article by the authors.~\cite{Jeon}

\begin{figure}
\centerline{\epsfig{file=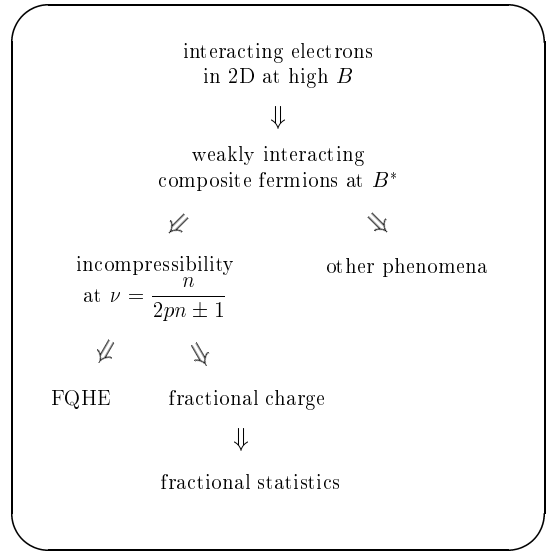,width=3.0in,angle=0}}
\caption{The logical path to fractional statistics.
First, the interacting electrons transform into weakly interacting 
composite fermions at an effective magnetic field.  Composite fermions 
form incompressible states at certain fractional fillings of the 
lowest Landau level.  Incompressibility implies fractional charge, which, 
finally, leads to fractional statistics.}  
\label{diagram}
\end{figure}

The logical route to fractional statistics is displayed in 
Fig.~\ref{diagram}, which serves to clarify the cause-and-effect relationship 
between various concepts.  The fractional statistics is a consequences of 
incompressibility at fractional filling factors;~\cite{Comment} the 
incompressibility itself results from the formation of composite fermions.  
Two notable facts consistent with the directions of arrows in 
Fig.~\ref{diagram} are that (a) the fractional statistics can be derived from 
composite fermions, but the reverse is not possible, and (b) the 
fractional statistics is tied to incompressibility, whereas composite 
fermions are more general and apply to compressible states as well. 
The CF theory is the microscopic theory of the QHS, whereas a  
description in terms of fractional statistics quasiparticles 
is an effective theory obtained when all but a few degrees of freedom are 
integrated out.

It should be stressed that fractional statistics is a rather  
delicate concept.  The effective magnetic field of composite fermions 
is a robust {\cal O}($N$) quantity, which has been directly measured 
in experiments, and gives an immediate explanation for the FQHE and 
many other phenomena.  
The fractional statistics, on the other hand, provides a 
natural interpretation for a subtle, but perhaps measurable,  
property of composite fermions, which specifies how the effective 
magnetic field changes upon an {\cal O}(1) variation in 
the particle density.  That is the reason why it has not been possible 
to confirm it so far, although several proposals have recently been 
advanced.~\cite{Proposals}

This paper is organized as follows:
Section II is devoted to the introduction of composite fermion
concept and the interpretation in terms of effective magnetic fields.
In Sec. III,
we will first calculate the Berry phase for a single CF quasiparticle, 
and show that it is consistent with the effective magnetic field principle.
The effective magnetic field principle is also shown to explain 
the situation when the CF quasiparticle lies outside the 
quantum Hall droplet.  Fractional statistics of CF quasiparticles 
is calculated microscopically in Sec. IV.
We will see that very small deviations in the trajectory make corrections 
that are on the same order as the statistics itself.
The CF quasiparticles in different CF-quasi-Landau levels 
are also found to exhibit the same fractional statistics.
Finally, constraints on experimental observations of the
fractional statistics are discussed.

\section{Composite fermions and effective magnetic field}

The physics of the QHS is governed by the Hamiltonian
\begin{equation}
H = \sum_{j}{1 \over 2m_b}\left [{\hbar \over i}{\vec\nabla}_{j}
+ {e \over c}{\vec{A}}(\vec{r}_j) \right ]^{2} +
\sum_{j < k} {e^2 \over |\vec{r}_j - \vec{r}_k|}  \;.
\end{equation}
In the limit of very large magnetic fields, all electrons 
can be taken to reside in the lowest Landau level
(LL), and the Hamiltonian reduces,
in appropriate units, to 
\begin{equation}
H={\cal P}_{LLL}\sum_{j<k}\frac{1}{|\vec{r}_j - \vec{r}_k|}{\cal P}_{LLL} \;,
\end{equation}
where ${\cal P}_{LLL}$ denotes projection into the lowest LL.

The problem is highly non-trivial because of the lack of a 
small parameter, but we have a good 
understanding of its physics, both qualitatively and quantitatively,
based on the formation of a new kind of fermions called composite fermions, 
which are bound states of electrons and an even number of quantized vortices.
Through composite fermions, many essential features can be understood through  
an analogy to the well understood integral quantum Hall effect (IQHE).
The wave function of the QHS has the following structure:
\begin{equation}
\Psi_{\nu}= {\cal P}_{LLL} \Phi_{\nu^*}\prod_{j<k}(z_j-z_k)^{2p}\;.
\label{cfwf}
\end{equation}
Here $\Psi_\nu$ is the wave function of
interacting electrons at filling factor $\nu$, 
defined as $\nu=\rho\phi_0/B$ ($\rho$ is the two-dimensional density of 
electrons, $B$ is the external magnetic field, and $\phi_0=hc/e$ is 
called the magnetic flux quantum).
$\Phi_{\nu^*}$ is the wave function for {\em weakly interacting} 
electrons at filling factor $\nu^*$, related to 
$\nu$ by 
\be
\nu=\frac{\nu^*}{2p\nu^* + 1}\;.
\ee
The complex number $z_j=x_j-i y_j$ denotes the position of the
$j$th electron in the $x$-$y$ plane.
$\Psi_\nu$ are known to be accurate representations of the actual eigenfunctions of 
the lowest LL projected Coulomb Hamiltonian,~\cite{Heinonen,JK,Dev} and 
it will be assumed below that they provide a correct account of 
topological properties like the fractional statistics.

The filling factor $\nu$ is typically $<1$, whereas $\nu^*$ is typically $>1$.
The mapping in Eq.~(\ref{cfwf}) leads to a simplification of the 
problem because the space of candidate wave functions at $\nu^*$  
is much smaller than that at $\nu$.  In particular, when $\nu^*=n$ (an integer),
the ground state wave function $\Phi_n$ is unique, giving 
a unique wave function $\Psi_\nu$ at $\nu=n/(2pn+1)$.
That explains the FQHE at these filling factors, which are the 
prominently observed sequences of fractions.

The physics of the wave function $\Psi_\nu$ is best understood 
prior to lowest LL projection, that is, with 
\begin{equation}
\Psi^{\rm up}_{\nu}= \Phi_{\nu^*}\prod_{j<k}(z_j-z_k)^{2p}\;.
\label{cfwfup}
\end{equation}
The Jastrow factor $\prod_{j<k}(z_j-z_k)^{2p}$
binds $2p$ vortices to each electron.  The bound state is interpreted
as a particle, namely the composite fermion. 
Now imagine a composite fermion, {\em i.e.} an electron along with $2p$ vortices,
traversing a closed loop enclosing an area $A$ (in the counterclockwise direction).
The phase associated with this process contains two terms:
\begin{equation}
\Phi^*=-2\pi\frac{BA}{\phi_0}+2\pi 2p  N_{\rm enc}
\label{phi*}
\end{equation}
where $N_{\rm enc}$ is the number of composite fermions inside the loop.
The first term is the usual Aharonov Bohm (AB) phase 
\begin{equation}
\Phi=-2\pi\frac{BA}{\phi_0}
\end{equation}
produced when a particle of charge $-e$ executes a counter clockwise loop, with 
the magnetic field pointing in the $+z$ direction.
The second term is the phase due to $2p$ vortices going 
around $N_{\rm enc}$ particles inside
the loop, with each particle contributing a phase of $2\pi$.
Equation~(\ref{phi*}) will play a fundamental role in what follows.
As we shall see, this equation 
implies incompressibility at certain fractional fillings,
and also explains the origin of fractional statistics.

We interpret the net phase as the AB phase due to an {\em effective}
magnetic field, $B^*$:
\begin{equation}
\Phi^* \equiv -2\pi\frac{B^* A}{\phi_0}\;.
\end{equation}
In a mean-field approximation, we 
replace $N_{\rm enc}$ by its average value $\rho A$, 
which gives
\begin{equation} \label{B*}
B^*=B-2p\rho\phi_0\;.
\end{equation}
Thus, the composite fermions behave as though they were in
a much smaller effective magnetic field.

To understand why the Berry phases coming from the vortices in the Jastrow 
factor effectively {\em cancel} (as opposed to augment) the magnetic field, 
it is instructive to understand the effective magnetic field 
by eliminating the phases of the Jastrow factor in favor of
a vector potential following the standard approach.~\cite{Wilczek,Girvin}
Consider the Schr\"odinger equation
\begin{eqnarray}
\lefteqn{\left[\frac{1}{2m_b}\sum_i\left(\vec{p}_i+\frac{e}{c}\vec{A}(\vec{r}_i)\right)^2
+V\right]  
\prod_{j<k}(z_j-z_k)^{2p}\Phi_{\nu^*}\nonumber}  \\
&&\hspace*{35mm} = E \prod_{j<k}(z_j-z_k)^{2p}\Phi_{\nu^*}
\end{eqnarray}
where $V$ is the interaction.  The kinetic energy term will be
the important one in what follows.
(We note that the unprojected wave function is not an exact eigenfunction 
of the Hamiltonian.  For the sake of the present argument, one may think of 
$\Phi_{\nu^*}$ as an arbitrary wave function rather than the solution 
of the non-interacting problem at $\nu^*$; then, the exact eigenstate 
in question can always be written in the above form.  While performing the actual 
calculations of the Berry phase, we will of course use the projected wave functions
which have a close to 100\% overlap with the exact eigenstates.) 
Display the phases due to the Jastrow factor explicitly:
\be
\prod_{j<k}(z_j-z_k)^{2p}=e^{-i 2p \sum_{j<k}\phi_{jk}}
\prod_{j<k}|z_j-z_k|^{2p} \;.
\ee
Here
\be
\phi_{jk}=i \ln \frac{z_j-z_k}{|z_j-z_k|} \;.
\ee
We have been careful above to keep track of the fact that
$z=re^{-i\phi}$, as appropriate for external magnetic field
in the $+z$ direction.
The Schr\"odinger equation can be rewritten as
\begin{widetext}
\begin{equation}
\left[\frac{1}{2m_b}\sum_i\left(\vec{p}_i+\frac{e}{c}\vec{A}(\vec{r}_i)
+ \frac{e}{c}\vec{a}(\vec{r}_i)\right)^2 +V\right]   
\prod_{j<k}|z_j-z_k|^{2p}\Phi_{\nu^*}\nonumber  
 =  E \prod_{j<k}|z_j-z_k|^{2p}\Phi_{\nu^*} \;,
\end{equation}
\end{widetext}
where the additional vector potential, that simulates the effect of
the phases of the Jastrow factor, is given by
\be
\vec{a}(\vec{r}_i)=-\frac{2p}{2\pi}\phi_0{\sum_j}^{'}\vec{\nabla}_i
\phi_{ij} \;,
\ee
where the prime denotes the condition $j\neq i$,
The corresponding magnetic field is
\be
\vec{b}_i= -2p \phi_0 \hat{z}
{\sum_j}^{'}\delta^2(\vec{r}_i-\vec{r}_j) \;.
\ee
Thus, the phase of the Jastrow factor is equivalent to each electron
seeing a flux tube of strength $-2p\phi_0$ on all other electrons;
the minus sign indicates that the flux tube points
in the $-z$ direction, opposite to the
direction to the external field $\vec{B}=B \hat{z}$.

This interpretation raises the following questions.
(i)  The effective vector potential does
not take care of {\em all} of the phases in the unprojected 
wave function $\Psi^{\rm up}_{\nu}$, because
there are additional vortices and anti-vortices in $\Phi_{\nu^*}$.
What about their effect?  (ii) How does the projection 
into the lowest LL affect the above considerations?  
The feature that $2p$ vortices are strictly bound to electrons 
prior to the projection is lost upon projection into the 
lowest LL.  For example, for $\nu>1/3$, where composite fermions
manifestly carry two vortices prior to projection,  
only one vortex can be bound to each electron {\em after} 
projection.  The projection thus obscures the 
physics of composite fermions.  Is there any way of seeing 
an effective magnetic field directly with the projected 
wave functions?

Even though the effective magnetic field 
is revealed most clearly in the unprojected 
wave functions, the robustness of the concept to perturbations  
has been confirmed in a model independent manner by numerous facts.
(i) Experiments clearly show a remarkably close correspondence 
between the FQHE and the IQHE, thus providing a non-trivial 
global confirmation of the effective magnetic field concept. 
(ii) Direct measurements of the cyclotron radius of the charge 
carrier~\cite{Goldman} are consistent with $B^*$.
(iii) Exact diagonalization studies show that the low energy 
spectrum of interacting electrons at $B$ has a one to one 
correspondence with the low energy spectrum of non-interacting 
fermions at $B^*$.~\cite{Dev2}
(iv) The wave functions of interacting electrons at $B$ ($\nu$)
are closely related to the wave functions of non-interacting 
electrons at $B^*$ ($\nu^*$), as seen in Eq.~(\ref{cfwf}).  From 
these observations, it is clear 
that the concept of effective magnetic field is more generally 
valid than the derivation based on the unprojected 
wave functions $\Psi^{\rm up}$ would suggest.

We now proceed to confirm Eq.~(\ref{phi*}) by 
calculating the Berry phase explicitly for a closed 
loop of composite fermion at $\nu=1/3$ and $\nu=2/5$ 
for the lowest LL projected wave functions.  The 
answers are fully consistent with the effective magnetic field 
principle.

\section{Berry phase for a single CF quasiparticle}

To confirm the effective magnetic field concept in a 
Berry phase calculation, one can envision creating a localized 
composite-fermion wave packet and determining the Berry phase associated 
with a closed loop enclosing an area $A$.  
Consider first the ground state at $\nu=n/(2pn+1)$, which maps 
into $\nu^*=n$ filled quasi-Landau levels of composite fermions 
at an effective magnetic field given by $B^*=B/(2pn\pm 1)$.  From the 
analogous case of $\nu=n$, where $n$ Landau 
levels are fully occupied, it is obvious that it is not possible to make a wave packet 
here without creating excitations.  
Therefore, one is forced to consider excitations.  
At $\nu=n$ we can straightforwardly make a wave packet 
if we put an additional electron in the lowest unoccupied LL, which can then be moved  
in any desired trajectory.  That is what we will do with composite fermions.

We will refer to as the ``composite-fermion quasiparticle" (CFQP) a 
composite fermion in the otherwise empty CF-quasi-LL,
which is the image of the electron state which has $n$ LLs completely occupied
and a single electron in the $(n+1)^{st}$ LL.  Similarly, the hole 
left behind when a composite fermion is removed
from the topmost CF-quasi-LL will be termed ``composite-fermion quasihole" (CFQH). 
The state at $\nu=n/(2pn\pm 1)$ is to be thought of as the ``vacuum."  
Relative to the ``vacuum" state, the CFQP or CFQH has a charge excess or deficiency in 
a spatially localized region.  It ought to be stressed that even a 
single CFQP or a CFQH describes a strongly correlated state of 
many interacting electrons.

Now take a CFQP in a loop enclosing an area $A$. 
Because it is nothing but a composite fermion, the phase is predicted to be 
the same as in Eq.~(\ref{phi*}):
\begin{equation} \label{eq:BP1}
\Phi^* = -2\pi \frac{B^*A}{\phi_0} = -2 \pi \frac{eBA}{(2pn+1)hc} \;.
\label{eq12}
\end{equation}
This is what we will first confirm.

The calculation of Berry phase requires microscopic wave functions
which are constructed starting with the wave function of a quasiparticle
at $\nu^*=n$, using the standard framework of the CF theory.  One problem 
is to figure out where to place the electrons in the 
corresponding IQHE problem, so, when the wave function 
is transformed to that of composite fermions, we get 
the CFQPs at the desired location.  
To do so, we implement the mapping 
from $\nu^*$ to $\nu$ {\em in a manner that preserves 
distances} (to zeroth order).  We first construct a quasiparticle
wave function at $B^*$, multiply it by $\Phi_1^{2p}$, where  
\be
\Phi_1=\prod_{j<k=1}^N(z_j-z_k)\exp\left[-\frac{1}{4l_1^2}\sum_i|z_i|^2\right]
\ee
with $l_1^2=\hbar c/eB_1=\hbar c/e \rho\phi_0$, 
and finally project the product into the lowest electronic LL.
This mapping preserves the size of the disk containing the quantum Hall 
droplet, because while the Jastrow factor pushes the 
particles out, the Gaussian pulls them in precisely by an amount to cancel 
the two effects.  It is easy to check that the density is not changed 
in going from $\nu^*=n$ to $\nu=n/(2pn+1)$ in this manner.
(See the article by Jain in Ref.~\onlinecite{Heinonen} for more details.)

At $\nu^*$, the single particle orbitals in the lowest LL are given by 
\be
\zeta_m(z) \equiv \frac{z^m}{\sqrt{2\pi 2^m m!}} 
\exp\left[ -\frac{1}{4l^{*2}}|z|^2\right]
\ee
where $l^*=(2pn+1)^{1/2}l$ is the magnetic length at  $B^*$.  
To put a CFQP at $\eta$, we first 
construct the electronic wave function at $\nu^*$ with an  
electron in the relevant Landau level (otherwise empty) at $\eta$
in a familiar coherent state.  The coherent state at $\eta$ in the lowest LL is 
given by 
\begin{eqnarray} \nonumber
\bar \phi^{(0)}_{\eta}(\vec{r}) &=& 
\sum_{m=0}^\infty \bar{\zeta}_m (\eta) \zeta_m (z)\\
&=&\exp\left[\frac{\bar{\eta}z}{2l^{*2}} 
-\frac{|\eta|^2}{4l^{*2}}-\frac{1}{4l^{*2}}|z|^2\right]\;.
\end{eqnarray}
One can elevate the coherent state to higher Landau levels by repeated application 
of the LL raising operator
$a^\dagger\equiv (2\partial/\partial z -
		\bar{z}/2)/\sqrt{2}$, which 
leads to the coherent-state wave function in the $(n+1)^{st}$ LL, apart
from a constant factor,
\be
\bar \phi^{(n)}_{\eta}(\vec{r})= (\bar{z}-\bar{\eta})^n 
\exp\left[\frac{\bar{\eta}z}{2l^{*2}}
-\frac{|\eta|^2}{4l^{*2}}-\frac{1}{4l^{*2}}|z|^2\right]\;.
\ee
It is convenient to define 
\be
\phi^{(n)}_{\eta}(\vec{r})= (\bar{z}-\bar{\eta})^n
\exp\left[\frac{\bar{\eta}z}{2l^{*2}} -\frac{|\eta|^2}{4l^{*2}}\right]
\ee
so 
\be
\bar \phi^{(n)}_{\eta}(\vec{r})= \phi^{(n)}_{\eta}(\vec{r}) 
\exp\left[ -\frac{1}{4l^{*2}}|z|^2\right]\;.
\ee

As an example, consider the system at $\nu=1/(2p+1)$, 
which is related to the CF filling $\nu^*=1$.
The electron wave function at $\nu^*=1$ with fully occupied lowest LL  
and an additional electron in the second LL at $\eta$ is  
\be
\Phi_1^{\eta}=\left|\begin{array}{ccccc}
\phi^{(1)}_{\eta}(\vec{r}_1) & \phi^{(1)}_{\eta}(\vec{r}_2) & . &.&.\\
1 & 1 & . &.&. \\
z_{1}&z_{2}&. &.&.\\
.&.&.&.&.\\
.&.&.&.&.  \\
z_1^{N-2} & z_2^{N-2} & . &.&.
\end{array}
\right|e^{-\sum_j|z_j|^2/4l^{*2}} \;.
\ee
This leads to the (unnormalized) wave function for a CFQP at $\nu=1/(2p+1)$:
\begin{eqnarray}
\Psi^{\eta}_{1/(2p+1)}\! &=&\! {\cal P}_{LLL} \!
\left|\begin{array}{ccc}
\phi^{(1)}_{\eta}(\vec{r}_1) & \phi^{(1)}_{\eta}(\vec{r}_2) & \ldots \\
1 & 1 & \ldots \\
z_{1}&z_{2}&\ldots\\
.&.&\ldots\\
.&.&\ldots\\
z_1^{N-2} & z_2^{N-2} & \ldots 
\end{array}
\right| \nonumber \\
& & \;\;\;\;\times \prod_{i<k=1}^N(z_i-z_k)^{2p}\! e^{-\sum_j|z_j|^2/4l^2}\;.
\label{1CFQP}
\end{eqnarray}
Here, we have used 
\be
\frac{1}{l^{*2}}+\frac{2p}{l_1^{2}}=\frac{1}{l^{2}}
\ee
which is equivalent to Eq.~(\ref{B*}). 
This wave function is similar to that considered 
by Kj{\o}nsberg and Leinaas,~\cite{Kjonsberg2} but not identical.

\begin{figure}
\centerline{\epsfig{file=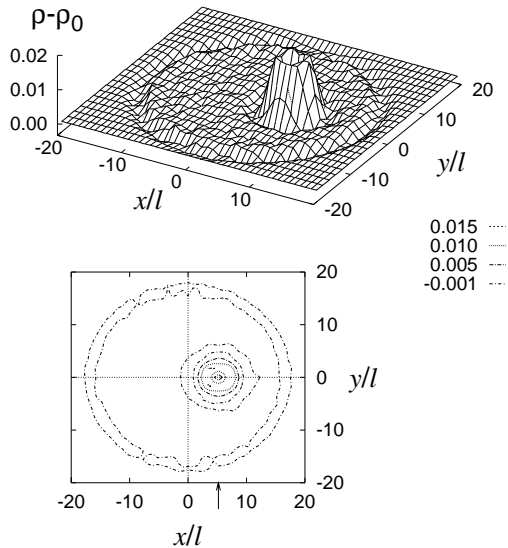,width=3.0in,angle=0}}
\caption{Excess charge density relative to the ground state 
in the presence of a single CFQP.
We used $\nu=1/3$, $N=50$, and 
$\eta/l=0.3R_d \approx 5.2$ with the disk size $R_d
\equiv\sqrt{2N/\nu}$.
The resulting position of the CFQP is in good agreement with 
the intended position, which
is indicated by the arrow in the contour plot (lower panel).}
\label{fig:ExcessDensity}
\end{figure}

Figure~\ref{fig:ExcessDensity} shows the excess density due to the
presence of a single localized CFQP for $\nu=1/3$.
The localized excess profile is clearly observed in the intended
position indicated by the arrow in the lower panel;
the profile has a smoke-ring-like shape since the quasiparticle is
in the second CF-quasi-LL.  (The coherent wave packet for an electron 
in the second LL also has a similar shape.) 
A deficit of the charge along the boundary is also discernible.

In a similar way we can construct the wave function for a CFQP of  
the state at $\nu=n/(2pn+1)$ for arbitrary $n$ and $p$.
For example, the wave function at $\nu=2/(4p+1)$ corresponding to $n=2$
is given explicitly by
\begin{eqnarray}
\Psi^{\eta}_{2/(4p+1)} &=& {\cal P}_{LLL} 
\left|\begin{array}{ccc}
\phi^{(2)}_{\eta}(\vec{r}_1) & \phi^{(2)}_{\eta}(\vec{r}_2) &\ldots\\
\bar{z}_1 & \bar{z}_2 & \ldots\\
\bar{z}_1 z_{1}&\bar{z}_2 z_{2}&\ldots\\
\vdots&\vdots&\ldots\\
\bar{z}_1 z_1^{N/2-2} & \bar{z}_2 z_2^{N/2-2} & \ldots \\
1 & 1 & \ldots\\
z_{1}&z_{2}&\ldots\\
\vdots&\vdots&\ldots \\
z_1^{N/2-1} & z_2^{N/2-1} & \ldots\end{array}
\right| \nonumber \\
& & \;\;\;\;\times  \prod_{i<k=1}^N(z_i-z_k)^{2p} e^{-\sum_j|z_j|^2/4l^2}\;.
\label{1CFQP2_5}
\end{eqnarray}

There are two methods for performing projection into the lowest LL.  
In one method,~\cite{Jach} (i) normal ordering the factor multiplying the 
gaussian factor  $e^{-\sum_j|z_j|^2/4l^2}$ by bringing all $\bar{z}_i$ to the left of 
$z_i$, and (ii) make the replacement $\bar{z}_i\rightarrow 2\partial/\partial z_i$ 
with the understanding that the partial derivatives do not act on the
Gaussian factor $e^{-\sum_j|z_j|^2/4l^2}$.
We employ a slightly different other projection method, 
described in Ref.~\onlinecite{JK}, which has many advantages in the numerical calculation,
especially for large systems.
In the CF theory, the unprojected wave function has the form:
\bea
\Psi^{\rm up} &=& 
\left|\begin{array}{ccccc}
\psi_1(z_1) & \psi_1(z_2) & . &.&.\\
\psi_2(z_1) & \psi_2(z_2) & . &.&.\\
.&.&.&.&.\\
.&.&.&.&. \\
\psi_N(z_1) & \psi_N(z_2) & . &.&.\\
\end{array}
\right| \nonumber \\
& & \;\;\;\;\times \prod_{i<k=1}^N(z_i-z_k)^{2p} e^{-\sum_j|z_j|^2/4l^2}\;.
\eea
Such wave functions can be rewritten in the form 
\be
\Psi^{\rm up} = e^{-\sum_j|z_j|^2/4l^2} 
\left|\begin{array}{ccccc}
\psi_1(z_1) J_1^p & \psi_1(z_2) J_2^p & . &.&.\\
\psi_2(z_1) J_1^p & \psi_2(z_2) J_2^p & . &.&.\\
.&.&.&.&.\\
.&.&.&.&. \\
\psi_N(z_1) J_1^p  & \psi_N(z_2) J_2^p & . &.&.\\
\end{array}
\right|
\ee
with $J_j \equiv \prod_{k\ne j} (z_j - z_k)$.
Then the projected wave function is given by projecting each element
of the determinant separately into the lowest Landau level by the 
method described above.

In order to test the robustness of the results to how the projection 
is carried out,  we have studied wave function projected by 
the two ways as well as the unprojected one for the CFQP at $\nu=1/3$. 
The results were independent of the employed state so long as
the position of the CFQP is far from the boundary of 
the system.

\subsection{Berry phase}

\begin{figure}
\centerline{\epsfig{file=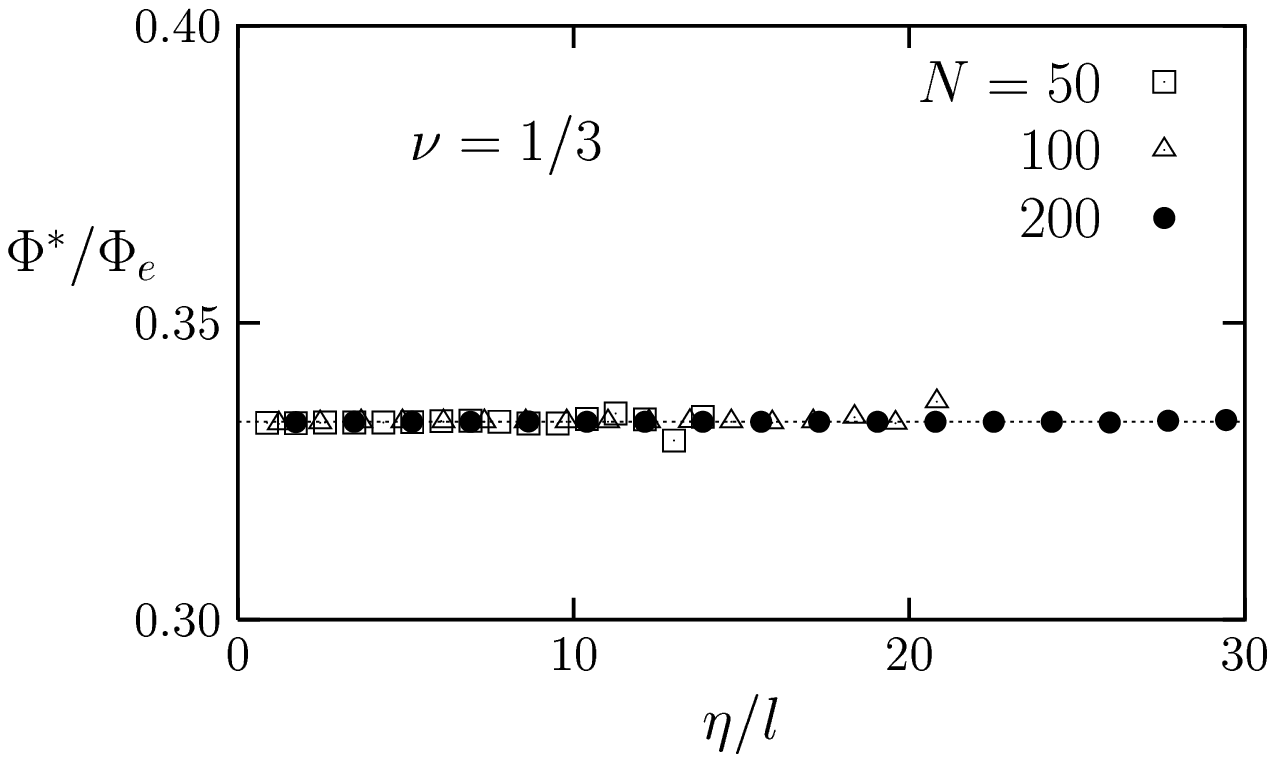,width=3.0in,angle=0}}
\centerline{\epsfig{figure=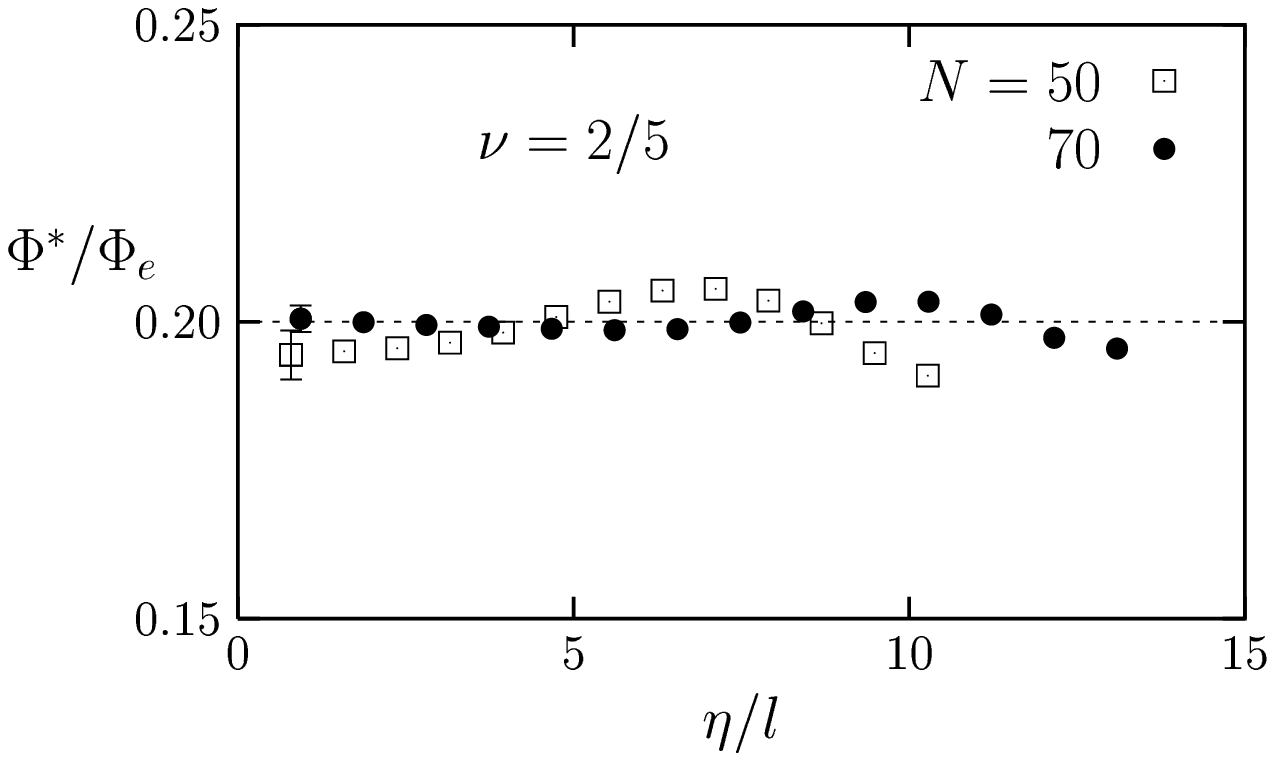,width=3.0in,angle=0}}
\caption{The Berry phase $\Phi^*$ for a single CFQP at 
$\nu=1/3$ (upper panel) and $\nu=2/5$ (lower panel) as a function of 
$\eta$.  
$N$ is the total number of composite fermions, $l$ is the 
magnetic length, and $\Phi_e\equiv -2\pi BA/\phi_0 $ is the Berry phase 
acquired by an electron moving in an empty space. 
The error bars from Monte Carlo sampling which are smaller than the 
symbol size are not shown explicitly.
The deviation at the largest $\eta/l$ for each $N$ is due to proximity to the edge.} 
\label{fig:BerryPhase}
\end{figure}

The Berry phase associated with a path ${\cal C}$ is given by
\be
\Phi^* =
\oint_{\cal C} d\theta \frac{\left<\Psi^\eta|
i\frac{d}{d\theta} \Psi^{\eta}\right>}{
\left<\Psi^\eta|\Psi^{\eta}\right>}\;,
\label{BP}
\ee
where $\Psi^\eta$ is the wave function containing a single CFQP at $\eta$.
For convenience, we take $\eta=R e^{-i\theta}$, and ${\cal C}$ refers to 
the circular path with $R$ fixed and $\theta$ varying from $0$ to $2\pi$ in the 
counterclockwise direction.  (Note that while the CFQP moves in the 
counterclockwise direction in the $x$-$y$ plane, the complex coordinate $\eta$ 
executes a clockwise loop in the complex plane.) 
The integrand in Eq.~(\ref{BP}) involves $2N$ dimensional integrals over 
the CF coordinates, which we evaluate by Monte Carlo method.  
Approximately $4 \times 10^{8}$ iterations are performed for each point.  
For $\nu=1/3$ we have studied systems with
$N=50$, 100, and 200 particles, and for $\nu=2/5$ we study systems with 
$N=50$ and 70 particles.  Projected wave functions are used in both cases.

The resulting values of Berry phase are displayed as a function of the
radius of the circular motion in Fig.~\ref{fig:BerryPhase}.
For both $\nu=1/3$ and $\nu=2/5$ the Berry
phase exhibits well-defined values, which agree well with those
in Eq.~(\ref{eq:BP1}) predicted by the effective magnetic field
description.  The deviation for large $\eta$ is a boundary effect, 
caused by the finiteness of the system.
The overall behavior for $\nu=1/3$ is consistent with the result in
Ref.~\onlinecite{Kjonsberg2}.  The effective magnetic field thus 
survives projection into lowest LL.

\subsection{Fractional local charge}

Above we derived the Berry phase as coming from the combination of two 
terms, due to an electron and $2p$ vortices going around a closed loop. 
This actually provides a derivation of the local charge of the CFQP, where the 
local charge, denoted by $-e^*$,
is defined to be charge excess associated with it relative 
to the uniform ground state.  The Berry phase of a CFQP is also 
the AB phase for a charge $-e^*$, which  
gives, using Eq.~(\ref{eq12}),
\be
-2\pi \frac{e^*BA}{hc}=-2\pi\frac{eBA}{(2pn+1)hc}\;.
\ee
Thus the local charge of a CFQP is 
\be
-e^*=-\frac{e}{2pn+1}\;.
\ee
The local charge can be derived in many other ways.~\cite{Laughlin}
Another way is to add the charge of the constituents of the CFQP, namely the 
electron and the vortices.~\cite{Heinonen}  The charge of a vortex~\cite{Laughlin} 
is $\nu e$, which is the 
occupation of a single orbital at filling $\nu$.  The local charge of
the CFQP, a bound state of an electron and $2p$ vortices,
is thus $-e^*=-e + 2p\nu e=-e/(2pn+1)$.
One can also show that the addition of one electron creates $2pn+1$ CFQPs,
which again implies that the local charge associated with each CFQP  
is $-e/(2pn+1)$.  The fact that the local charge is independent of 
details (relying only on incompressibility) provides insight into 
why the Berry phase of the CFQP is robust to projection into 
the lowest LL.

\subsection{CF quasiparticle outside the disk}

In the previous sections, we have considered only the situation when  
the CFQP is inside a QHS droplet.  It is interesting to ask what 
happens when a CFQP is located outside the droplet.
Far from the droplet, the CFQP no longer has any other CFs nearby and
therefore there is no screening hole.  
Its local charge therefore is the same as a bare
charge $-e$ due to the absence of the screening cloud.
How about the Berry phase acquired by the CFQP?
Should it be the same as $\Phi_e$, which is the Berry phase for an electron
moving in the free space in a uniform external magnetic field $B$?

\begin{figure}
\centerline{\epsfig{file=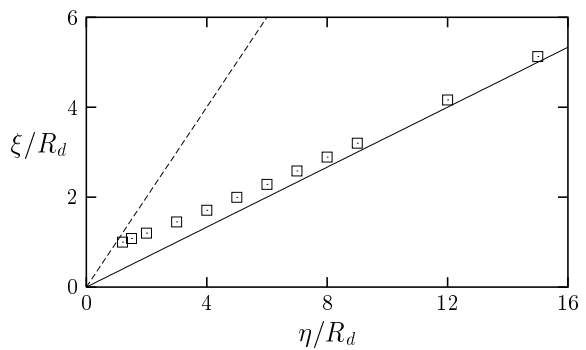,width=3.0in,angle=0}}
\caption{Actual location $\xi$ of a CFQP (in units of the disk size 
$R_d \equiv l \sqrt{2N/\nu} $) as a function of the parameter
$\eta$ for $\nu=1/3$ and $N=50$ when the CFQP is outside the quantum Hall
superfluid droplet.
The dashed line is $\xi=\eta$ and the solid line is $\xi=\eta/(2pn+1)$, 
the position the CFQP would have 
in the absence of any other composite fermions (in the case of $\nu=1/3$, we 
have $\xi=\eta/3$; see text).
}
\label{fig:OutsidePosition}
\end{figure}

Before proceeding further within the CF theory, we should re-examine  
how the actual CFQP position is related to the parameter $\eta$ in the 
wave function.
The condition that the position of the CFQP is given by $\eta$ 
was derived under the assumption that the
CFQP is surrounded by other uniform CFs;  we can no longer expect that
the CFQP position is given by $\eta$ when it is off the droplet.
Far from the droplet, the CFQP will experience the {\em bare} 
external magnetic field $B$ rather than $B^*=B/(2pn+1)$.
Accordingly, the usage of effective magnetic length $l^*$ 
in the coherent state leads to the actual position of the CFQP 
given by $\xi \approx \eta/(2pn+1)$.
This is verified in Fig.~\ref{fig:OutsidePosition}, which plots 
the actual location $\xi$ obtained numerically as a function of the
parameter $\eta$. 
The numerical calculation was performed at $\nu=1/3$ for $N=50$
composite fermions.
When the parameter $\eta$ exceeds the droplet size $R_d \equiv l
\sqrt{2N/\nu}$, the  position $\xi$ deviates from the (dashed) line
$\xi=\eta$.
As $\eta$ is increased further, $\xi$ approaches the (solid) line 
$\xi=\eta/(2pn+1)$.

The CF theory also makes a prediction for the Berry phase of a single
CFQP outside the droplet.
Since the enclosed area is not filled uniformly with CFs, we can no longer
use the {\em uniform} effective magnetic field in Eq.~(\ref{B*}).
Instead, we must use Eq.~(\ref{phi*}). 
Outside the droplet, the number of enclosed composite fermions
is $N_{\rm enc}=N-1$
and the enclosed area is $A=\pi \xi^2$,
yielding the Berry phase
\be
\frac{\Phi^*}{\Phi_e} = 1 - \frac{4p(N-1)}{(\xi/l)^2}.
\label{eq:BPoutside}
\ee
The second term, the contribution from the composite fermions on the 
QHS droplet, is of 
order ${\cal O}(\xi^{-2})$.

\begin{figure}
\centerline{\epsfig{file=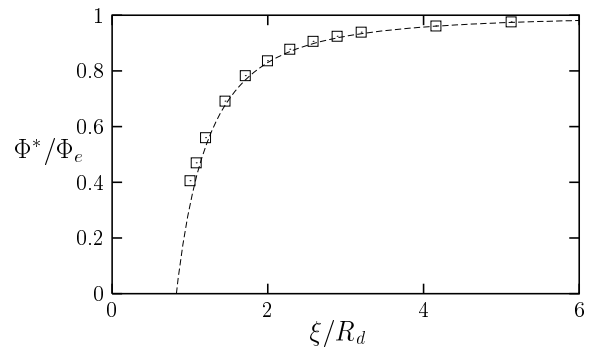,width=3.0in,angle=0}}
\caption{Berry phase $\Phi^*$ acquired by a CFQP 
when it is outside the quantum Hall droplet.
The system has $N=50$ composite fermions at the filling $\nu=1/3$. 
The dashed line is the prediction of the CF theory, 
and the squares are calculated from the microscopic wave functions.
}
\label{fig:OutsideBP}
\end{figure}

Figure~\ref{fig:OutsideBP} demonstrates clearly that 
the prediction in Eq.~(\ref{eq:BPoutside}), 
denoted by the dashed line in the figure,
explains nicely the behavior of $\Phi^*$ when the CFQP is
outside the droplet.
Indeed, the numerical data begin to deviate when the CFQP approaches
the boundary of the droplet, and eventually give a definite value
$1/(2pn+1)$ inside the system (the Berry phase for a CFQP 
inside the disk are not shown in Fig.~\ref{fig:OutsideBP}).
Because the {\em local} charge of a CFQP becomes $-e$
just outside the droplet, the long tail $\sim {\cal O}(\xi^{-2})$ in
$\Phi^*/\Phi_e$ is not explained by the alternate interpretation 
of the Berry phase in terms of a charge of $-e^*$ moving under 
the external magnetic field $B$.

\section{Two CF quasiparticles: fractional statistics}

We have confirmed Eq.~(\ref{phi*}) for a single CFQP
in a closed loop, when the other composite fermions make a uniform
background state.  How about the situation when the density is 
{\em not} uniform?  The simplest question one may ask 
is: How does the Berry phase change when we change the number of enclosed
particles in a loop by a number of order unity? 
Following Ref.~\onlinecite{Goldhaber}, Equation~(\ref{phi*}) predicts  
\begin{equation}
\Delta\Phi^*=2\pi 2p \Delta \langle N_{\rm enc}\rangle \;.
\end{equation}
To be specific, we will add a single CFQP inside the loop,
which, counting the correlation hole around it, carries an excess of 
$\Delta \langle N_{\rm enc}\rangle=1/(2pn+1)$ electrons, which gives:
\be
\Delta\Phi^*=2\pi \frac{2p}{2pn+1}\equiv 2\pi \theta^*
\label{eq:DeltaPhi}
\ee
with 
\be
\theta^*=\frac{2p}{2pn+1}\;.
\label{theta*}
\ee
A fractional value of $\theta^*$ is often interpreted through 
an assignment of a fractional statistics to the CFQPs.  
Note that the fractionally quantized value for $\theta^*$ is a 
direct consequence of the fractional quantization of the local charge.
It should also be stressed that $\theta^*$ is a much more subtle quantity 
than $B^*$, sensitive to order unity changes in the enclosed 
particle number.  Equation~(\ref{phi*}) is surely correct in a mean-field 
sense, but it is by no means obvious that it captures O(1) effects  
accurately.

The meaning of fractional statistics is complicated in the 
QHS context by the presence of a magnetic 
field, which produces its own phase for any closed loop, even 
when it does not include another CFQP. 
(Of course {\em all} loops enclose other composite fermions; here
we think of only the excitations as the CFQPs.)
The fractional statistics is 
defined as the {\em difference} in the phase for a given closed 
path when one CFQP is added to the interior. 
It is a small perturbation on a large effect. 
Even though we have derived the fractional statistics as an immediate 
corollary of the effective magnetic field principle,~\cite{Goldhaber}
the value of $\theta^*$ had been derived prior to the CF theory 
from general arguments,~\cite{Halperin,Su} assuming incompressibility 
at a fractional filling; the earlier values (if evaluated with $\vec{B}$
in the $+z$ direction) differ from 
the one quoted here by 1 ($mod$ 2).  The reason for this
deviation is that our theory deals with quasiparticles that
obey fermionic exchange statistics, so an additional factor (-1)
arises from the mere position exchange of two CF quasiparticles in the
microscopic wave function,
whereas the previous ones assume bosonic exchange statistics.
In Ref.~\onlinecite{Kjonsberg2}  the result was shifted by unity.
We prefer not to do that.  The quasiparticles that we work with are the
actual quasiparticles (as would be obtained in an exact diagonalization
study if, say, two impurities were placed to localize two quasiparticles),
and therefore the phases given below are what an
actual experiment would measure.

Our goal is to confirm Eq.~(\ref{theta*}) in a microscopic calculation.
The statistics parameter is given by
\be
\theta^* =
\oint_{\cal C} \frac{d\theta}{2\pi} \frac{\left<\Psi^{\eta,\eta'}| i\frac{d}{d\theta}
\Psi^{\eta,\eta'}\right>} {\left<\Psi^{\eta,\eta'}|\Psi^{\eta,\eta'}\right>}-
\oint_{\cal C} \frac{d\theta}{2\pi}
\frac{\left<\Psi^\eta|
i\frac{d}{d\theta} \Psi^{\eta}\right>}{
\left<\Psi^\eta|\Psi^{\eta}\right>}\;,
\label{Berry}
\ee
where $\Psi^\eta$ is the wave function containing a single CFQP at $\eta$, and 
$\Psi^{\eta,\eta'}$ has two CFQPs at $\eta$ and $\eta'$.
Here we take $\eta=R e^{-i\theta}$, and ${\cal C}$ refers to the path 
with $R$ fixed and $\theta$ varying from $0$ to $2\pi$ 
in the counterclockwise direction, as in the calculation of a single
CFQP Berry phase.
For convenience, we will take $\eta'=0$ and denote the microscopic
numerical value of $\theta^*$ by $\tilde{\theta}^*$(the reason will be
clear below).

\begin{figure}
\centerline{\epsfig{file=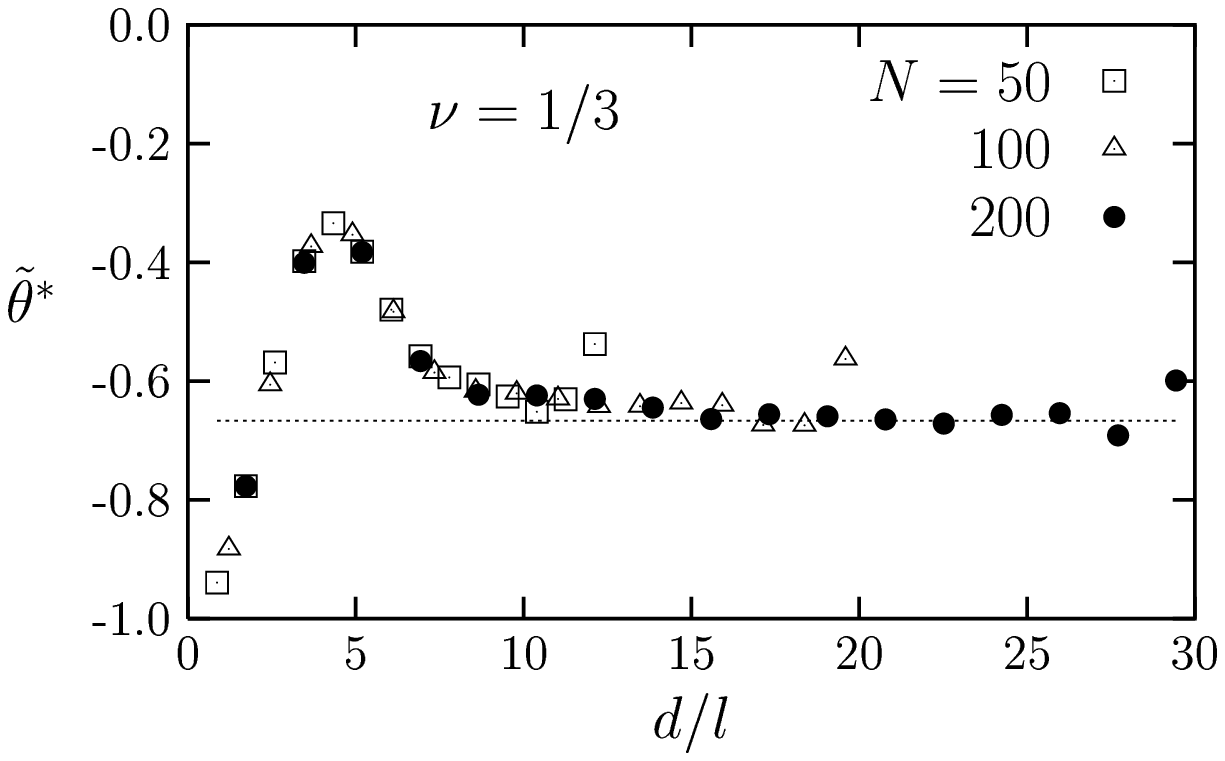,width=3.0in,angle=0}}
\centerline{\epsfig{figure=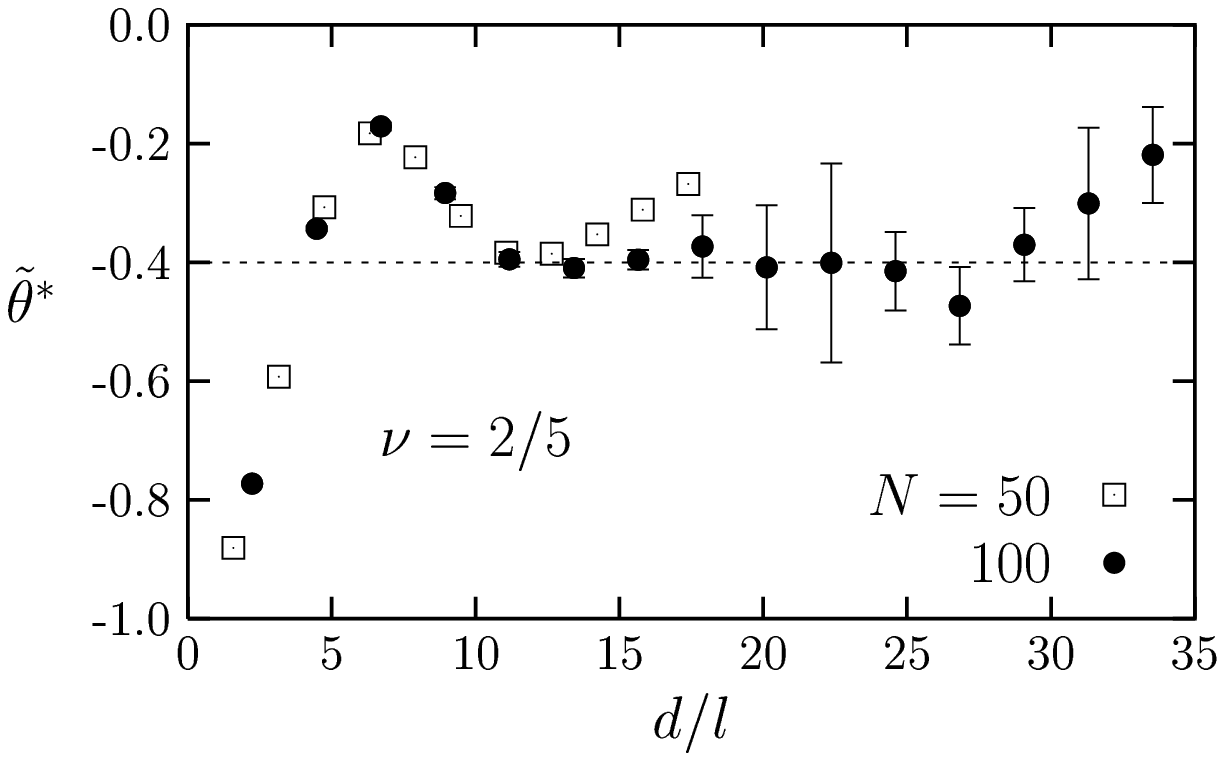,width=3.0in,angle=0}}
\caption{The statistical angle $\tilde\theta^*$ for the CFQPs at 
$\nu=1/3$ (upper panel) and $\nu=2/5$ (lower panel) as a function of 
$d \equiv |\eta-\eta'|$.  $N$ is the total number of composite fermions, and $l$ is the 
magnetic length. The error bar from 
Monte Carlo sampling is not shown explicitly when it is smaller than the symbol size.
The deviation at the largest $d/l$ for each $N$ is due to proximity to the edge.
This figure was shown earlier in Ref.~\protect\onlinecite{Jeon}, and is  
reproduced here for completeness.} 
\label{fig:stat}
\end{figure}

Before proceeding further, we mention a curious 
fact to illustrate the fragility of fractional statistics. 
Laughlin had proposed the following wave function for two 
quasiparticles at $\nu=1/m$ ($m$ odd):
\begin{eqnarray} \nonumber
\Psi_L^{\eta} &=& e^{-\sum_j|z_j|^2/4l^2}
\prod_{j=1}^N (\partial_{z_j} - \eta)
\prod_{j<k=1}^N (z_j - z_k)^m,\\ \nonumber
\Psi_L^{\eta,\eta'} &=& e^{-\sum_j|z_j|^2/4l^2}\\
&&\times \prod_{j=1}^N (\partial_{z_j} - \eta)(\partial_{z_j} - \eta')
\prod_{j<k=1}^N (z_j - z_k)^m .
\label{L2q}
\end{eqnarray}
Kj{\o}nsberg and Myrheim~\cite{Kjonsberg1} found 
that the Berry phase calculation of the statistics of the 
quasiparticle using this wave 
function does not produce a well defined answer.  
It was realized by Kj{\o}nsberg and Leinaas~\cite{Kjonsberg2} that 
the more accurate wave function of the CF theory 
produces a well defined value for $\theta^*$.  
What makes it all the more surprising is that both the 
wave functions of Laughlin and Jain produce the correct local charge 
for a single quasiparticle.  It is not understood why one of them fails 
to produce proper statistics, but the example 
underscores how the statistics may be sensitive to rather subtle correlations 
in the wave function.

In the CF theory,  the wave function for two CFQPs at $\nu=1/(2p+1)$
is a natural extension of that containing a single CFQP.
The electron wave function at $\nu^*=1$ with fully occupied lowest LL  
and two additional electrons in the second LL at $\eta$ and $\eta'$ is  
\be
\Phi_1^{\eta,\eta'}=\left|\begin{array}{ccccc}
\phi^{(1)}_{\eta}(\vec{r}_1) & \phi^{(1)}_{\eta}(\vec{r}_2) & . &.&.\\
\phi^{(1)}_{\eta'}(\vec{r}_1) & \phi^{(1)}_{\eta'}(\vec{r}_2) & . &.&.\\
1 & 1 & . &.&. \\
z_{1}&z_{2}&. &.&.\\
.&.&.&.&.\\
.&.&.&.&.  \\
z_1^{N-3} & z_2^{N-3} & . &.&.
\end{array}
\right|e^{-\sum_j|z_j|^2/4l^{*2}} \;.
\ee
This leads to the (unnormalized) wave function for two CFQPs at $\nu=1/(2p+1)$:
\begin{eqnarray}
\Psi^{\eta,\eta'}_{1/(2p+1)}\!\! &=&\!\! {\cal P}_{LLL}\!\! 
\left|\begin{array}{ccc}
\phi^{(1)}_{\eta}(\vec{r}_1) & \phi^{(1)}_{\eta}(\vec{r}_2) & \ldots \\
\phi^{(1)}_{\eta'}(\vec{r}_1) & \phi^{(1)}_{\eta'}(\vec{r}_2) & \ldots \\
1 & 1 & \ldots \\
z_{1}&z_{2}&\ldots \\
.&.&\ldots\\
.&.&\ldots \\
z_1^{N-3} & z_2^{N-3} & \ldots
\end{array}
\right| \nonumber \\
& & \;\;\;\;\times  \prod_{i<k=1}^N
\!\!(z_i-z_k)^{2p} \!
 e^{-\sum_j|z_j|^2/4l^2}\;.
\label{2CFQP}
\end{eqnarray}
The extension to the general filling $\nu=n/(2pn+1)$ is
again straightforward.
For reference, we give an explicit expression of the two CFQPs
wave function at $\nu=2/5$:
\begin{eqnarray}
\Psi^{\eta,\eta'}_{2/5}\! &=&\! {\cal P}_{LLL}\! 
\left|\begin{array}{ccc}
\phi^{(2)}_{\eta}(\vec{r}_1) & \phi^{(2)}_{\eta}(\vec{r}_2) &\ldots\\
\phi^{(2)}_{\eta'}(\vec{r}_1) & \phi^{(2)}_{\eta'}(\vec{r}_2) &\ldots\\
\bar{z}_1 & \bar{z}_2 &\ldots \\
\bar{z}_1 z_{1}&\bar{z}_2 z_{2}&\ldots\\
\vdots&\vdots&\ldots\\
\bar{z}_1 z_1^{N/2-3} & \bar{z}_2 z_2^{N/2-3} & \ldots \\
1 & 1 &\ldots \\
z_{1}&z_{2}&\ldots\\
\vdots&\vdots&\ldots\\
z_1^{N/2-1} & z_2^{N/2-1} & \ldots
\end{array}
\right| \nonumber \\
& & \;\;\;\;\times \prod_{i<k=1}^N
\!(z_i-z_k)^{2} \!
 e^{-\sum_j|z_j|^2/4l^2}\;.
\label{2CFQP2_5}
\end{eqnarray}

The statistics parameter $\tilde{\theta}^*$ for $\nu=1/3$ and $\nu=2/5$
was shown in Ref.~\onlinecite{Jeon}, reproduced in Fig.~\ref{fig:stat} for 
completeness.  $\tilde{\theta}^*$ takes a well-defined value for large
separations.  
At $\nu=1/3$ it approaches the asymptotic value of $\tilde{\theta}^*=-2/3$,
which is consistent with that obtained in Ref.~\onlinecite{Kjonsberg2}
without lowest LL projection.  
The calculation at $\nu=1/3$ explicitly demonstrates
that $\tilde{\theta}^*$ is independent of whether the projected or the
unprojected wave function is used.
Assuming the same is true for other fractions, we have performed 
the calculation at $\nu=2/5$ without the projection.  (The calculation 
of the statistics, a small difference between two large quantities,  
requires much greater accuracy than the calculation of $B^*$ considered in 
the previous section.  The use of projected wave functions is in principle 
possible, but very costly in terms of computation time.)
At $\nu=2/5$ the system size is 
smaller and the statistical uncertainty bigger, but the asymptotic value 
is clearly seen to be $\tilde{\theta}^*=-2/5$.  
At short separations there are 
substantial deviations in $\tilde{\theta}^*$; it reaches the asymptotic 
value only after the two CFQPs are separated by more than  
$\sim$ 10 magnetic lengths.  
Such deviations are presumably due to a significant overlap between
CFQPs when they are close.
(In contrast, the effective magnetic field is well defined for 
arbitrarily small closed loops.)

\subsection{The sign puzzle}

The microscopic value $\tilde{\theta}^*$ 
obtained above has the same magnitude as 
$\theta^*$ in Eq.~(\ref{theta*}) {\em but the opposite sign}. 
The sign discrepancy, if real, cannot be 
reconciled with Eq.~(\ref{phi*}) and would cast doubt on the 
fundamental interpretation of the CF physics in terms of an effective 
magnetic field.

To gain insight into the issue,  
consider two composite fermions in the otherwise empty lowest LL, 
for which various quantities can be obtained analytically. 
When there is only one composite fermion at $\eta=Re^{-i\theta}$, 
it is the same as an electron, with the wave function given by
\be
\chi^{\eta}=\exp\left[\frac{\bar{\eta}z}{2l^2}
-\frac{R^2}{4l^2}-\frac{|z|^2}{4l^2}\right]\;.
\ee
For a closed loop, 
\be
\oint_{\cal C} \frac{d\theta}{2\pi}
\frac{\left<\chi^\eta|
i\frac{d}{d\theta} \chi^{\eta}\right>}{
\left<\chi^\eta|\chi^{\eta}\right>}=-\frac{R^2}{2l^2}=-\frac{\pi R^2B}{\phi_0}\;.
\ee
Two composite fermions, one at $\eta$ and the other at 
$\eta'=0$, are described by the wave function 
\be
\chi^{\eta,0}=(z_1-z_2)^{2p}(e^{\bar\eta z_1/2}-e^{\bar\eta z_2/2})
e^{-(R^2+|z_1|^2+|z_2|^2)/4}.
\label{eq:2CFLL}
\ee
Here, we expect $\theta^*=2p$. 
However, an explicit evaluation of the Berry phase shows, 
neglecting {\cal O}($R^{-2}$) terms 
\be
\oint_{\cal C} \frac{d\theta}{2\pi}
\frac{\left<\chi^{\eta,0}|
i\frac{d}{d\theta} \chi^{\eta,0}\right>}{
\left<\chi^{\eta,0}|\chi^{\eta,0}\right>}=-\frac{R^2}{2l^2}-2p\;,
\label{wrong}
\ee
which gives $\tilde{\theta}^*=-2p$ for large $R$.  Again, it apparently has 
the ``wrong" sign.

A calculation of the density for $\chi^{\eta,0}$ shows that 
the actual position of the outer composite fermion is not $R=|\eta|$ but 
$R'$, given by
\be
\frac{R'^2}{l^2}=\frac{R^2}{l^2}+4\cdot 2p
\label{R2}
\ee
for large $R$.  This can also be seen in the inset of
Fig.~2 of Ref.~\onlinecite{Jeon}.
The correct interpretation of Eq.~(\ref{wrong}) therefore is 
\be
\oint_{\cal C} \frac{d\theta}{2\pi}
\frac{\left<\chi^{\eta,0}|
i\frac{d}{d\theta} \chi^{\eta,0}\right>}{
\left<\chi^{\eta,0}|\chi^{\eta,0}\right>}=-\frac{R'^2}{2l^2}+2p ,
\ee
which produces $\theta^*=2p$.  The {\cal O}(1) correction to the area enclosed 
thus makes a non-vanishing correction to the statistics.
(It is noted that the CFQP at $\eta=0$ is also a little off center, 
and executes a tiny circular loop which provides another correction to the 
phase, but this contribution vanishes in the limit of large $R$.)

This exercise tells us that an implicit assumption made in the earlier analysis,
namely that the position of the outer CFQP labeled by $\eta$ remains 
unperturbed by the insertion of another CFQP, leads to an incorrect value 
for $\theta^*$.  In reality, inserting another CFQP inside the loop pushes 
the CFQP at $\eta$ very slightly outward.

To determine the correction at $\nu=n/(2pn+1)$, we note that the 
mapping into composite fermions preserves distances to zeroth order,
so Eq.~(\ref{R2}) ought to be valid also at $\nu=n/(2pn+1)$.
This is consistent with the shift seen in Fig.~2 of Ref.~\onlinecite{Jeon}
for the position of the CFQP calculated numerically directly from the 
wave function.  Our earlier result
\be
\oint_{\cal C} \frac{d\theta}{2\pi}
\frac{\left<\Psi^{\eta,0}|
i\frac{d}{d\theta} \Psi^{\eta,0}\right>}{
\left<\Psi^{\eta,0}|\Psi^{\eta,0}\right>}=-\frac{R^2}{2l^{*2}}-\frac{2p}{2pn+1}
\ee
ought to be rewritten, using $l^{*2}/l^2=B/B^*=2pn+1$, as  
\be
\oint_{\cal C} \frac{d\theta}{2\pi}
\frac{\left<\Psi^{\eta,0}|
i\frac{d}{d\theta} \Psi^{\eta,0}\right>}{
\left<\Psi^{\eta,0}|\Psi^{\eta,0}\right>}=-\frac{R'^2}{2l^{*2}}+\frac{2p}{2pn+1}.
\ee
When the contribution from the closed path without the other CFQP, 
$-R'^2/2l^{*2}$, is subtracted out, $\theta^*$ of Eq.~(\ref{theta*}) is obtained.
The neglect of the correction in the radius of the loop introduces an error 
which just happens to be twice the negative of the ``correct" answer.

Before ending this subsection we note another subtle effect.
A quasiparticle in the bulk induces 
a quasihole at the boundary, the charge of which is non-uniformly distributed over the 
edge when the bulk quasiparticle is off-center.  As the primary quasiparticle is taken around 
a loop, the ``center" of the induced edge quasihole also executes a complete loop.  The 
contribution of the latter to the Berry phase is neglected in the heuristic 
derivation of the statistics as well as in the analytical calculation of Arovas 
{\em et al}.~\cite{Arovas}, but is explicitly included in the numerical calculations with a 
boundary.  The consistency of the numerical results with the heuristic expectation 
indicates that the boundary effects are negligible, at least so long as the 
primary quasiparticles are sufficiently far from the edge.

\subsection{Approach to the asymptotic value}

In the previous section, it was shown that the asymptotic value of the 
statistic parameter is explained within the CF theory.
The next question is how the asymptotic value is reached 
as the distance between two CFQPs is increased.
In Fig~\ref{fig:stat}, particularly for $\nu=1/3$, we can see that
$\tilde{\theta}^*$ approaches its asymptotic value very slowly 
even for $d\gtrsim 10 l$.  Is that slow convergence real, or only 
a result of the fact that the actual position of the CFQP has slight 
corrections?  Should the slow convergence persist for $\theta^*$, 
that would cast doubt on the usefulness of the concept of 
fractional statistics.

\begin{figure}
\centerline{\epsfig{file=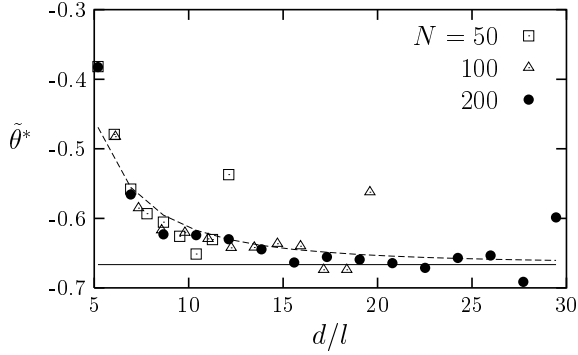,width=3.0in,angle=0}}
\caption{The statistical angle $\tilde\theta^*$ for the CFQPs at 
$\nu=1/3$ for large $d \equiv |\eta-\eta'|$.  
$N$ is the total number of composite fermions, and $l$ is the magnetic length. 
The dashed line is Eq.~(\ref{eq:correction_n}), which is in good
agreement with the long tail of the numerical data.  
The points near the edge deviate significantly from the dashed line.
}
\label{fig:further}
\end{figure}

To examine the origin of such long tail, we consider in
more detail two composite fermions in the lowest CF-quasi-LL.
For $^2$CF ($p=1$), 
we can explicitly calculate the statistics parameter in
Eq.~(\ref{Berry}) through the use of the wave function 
in Eq.~(\ref{eq:2CFLL}), leading to
\be
\tilde{\theta}^* =  \frac{
\displaystyle -\frac{R^2}{2l^2} \left[
   \frac{4R^2}{l^2}+ 32 + e^{ -R^2/2l^2} \left(
   \frac{R^4}{l^4} - \frac{20R^2}{l^2} +64 
   \right) \right] }{
\displaystyle
	\left[
	\frac{R^4}{l^4}+\frac{16R^2}{l^2} +32
   -e^{ -R^2/2l^2} \left(\frac{R^4}{l^4}-\frac{16R^2}{l^2} +32
   \right)\right].  }
\label{eq:CFLLtheta}
\ee
In the limit of $R \gg l$, $\tilde{\theta}^*$ reduces to
\be
\tilde{\theta}^* = - 2 + 16 \left(\frac l R \right)^2 
+ {\cal O}\left(\frac l R \right)^4 .
\ee
As observed for $\nu=1/3$,
we find that the deviation of $\tilde{\theta}^*$ from the asymptotic 
value decays only algebraically.

The density profile for two $^2$CFs on the lowest CF-quasi-LL 
is straightforwardly computed to be
\bea \nonumber
\rho^{\eta,0}(x) &\propto& e^{-(x-R)^2/2l^2} (x^4 + 8x^2 + 8)\\ \nonumber
&-&2 e^{-(R^2-Rx+x^2)/2l^2} [ x^2(x{-}R)^2{+}8 x(x{-}R){+}8]\\
&+& e^{-x^2/2} [ (x-R)^4 + 8(x-R)^2 +8]
\eea
along the $x$-axis, with the outer composite fermion intended to be
located at $(R,0)$.
As discussed in the previous section,
the actual positions $R'$ of the outer composite fermion is given by 
\bea \nonumber
\frac{R'}{l} &=& \frac R l + 4\left(\frac l R \right) - 32 \left(\frac l R\right)^3
+ {\cal O}\left(\frac l R \right)^5 \\
	&\equiv& R + \Delta R.
\eea
At the same time, the inner composite fermion also shifts to $(R'',0)=(-\Delta R,0)$. 
The Berry phase (divided by $2\pi$) 
acquired due to the position shift of the composite fermions is 
\bea \nonumber
\Delta \theta^* &=& - \frac{B \Delta A}{\phi_0} \\ \nonumber
&=& - \frac{1}{2\pi l^2} \left[ \pi  R'^2 + \pi R''^2 
- \pi R^2 \right]\\
  &=& - 4 + 16 \left(\frac l R \right)^2 
+ {\cal O}\left(\frac l R \right)^4 .
\eea
The real statistical parameter, $\theta^*=\tilde{\theta}^*-\Delta \theta^*$,
is given by 
\be
\theta^*=2+{\cal O}\left(\frac l R \right)^4
\ee
with the $+{\cal O}\left(\frac l R \right)^2$ term canceling out.
Thus, the power law tail in the difference between the CF value  
$\theta^*=2$ and the microscopic value $\tilde{\theta}^*$ in
Eq.~(\ref{eq:CFLLtheta}) is not real, but caused by a shift in the 
positions of the CFQPs.

If the same argument holds for nonzero $n$ and $p=1$, 
the additional Berry phase
(divided by $2\pi$) due to the position shift can be written as
\be
\Delta \theta^* =
- \frac4{2n+1} + \frac{16}{2n+1} \left(\frac l d \right)^2 
+ {\cal O}\left(\frac l d \right)^4 
\ee
through the use of the effective magnetic field $B^*/B = 1/(2n+1)$.
Adding the asymptotic value $\theta^*=2/(2n+1)$ gives
\be
\tilde{\theta}^* =
-\frac2{2n+1} + \frac{16}{2n+1} \left(\frac l d \right)^2 
+ {\cal O}\left(\frac l d \right)^4 \;.
\label{eq:correction_n}
\ee
This heuristic prediction of the CF theory 
is plotted in Fig.~\ref{fig:further}
(dashed line), and agrees well with the 
long tail of the numerical behavior for large $d/l$.

\subsection{Two nearby CF quasiparticles}

We now turn to the situation when the two CFQPs are located very close to
one another.  When the distance becomes comparable to the size of the CFQPs, it  
is not possible to define the distance between the CFQPs in 
a meaningful manner, so we will consider here the dependence of 
fractional statistics on $d=|\eta-\eta'|$,
which is a parameter entering the wave function.

\begin{figure}
\centerline{\epsfig{file=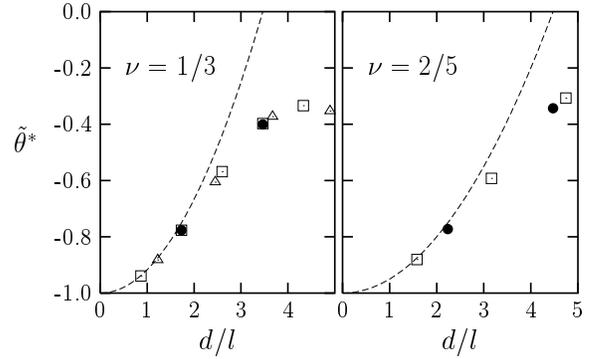,width=3.0in,angle=0}}
\caption{The statistical angle $\tilde\theta^*$ for the CFQPs at 
$\nu=1/3$ (left panel) and $\nu=2/5$ (right panel) 
for small $d \equiv |\eta-\eta'|$.  
The symbols are the same as in Fig.~\ref{fig:further}, 
and $l$ is the magnetic length. 
The heuristic formula in Eq.~(\ref{eq:near}) (dashed lines) 
agrees well with the actual result for small $d$.  }
\label{fig:near}
\end{figure}

The microscopic value $\tilde{\theta}^*$ for small $d$, as shown 
in Fig.~\ref{fig:near}, exhibits significant deviation from its
asymptotic value.  For very small $d$ it grows monotonically from $-1$
before undergoing a crossover to the asymptotic value.
To gain insight into this behavior, we again resort to CFs in an
otherwise empty lowest CF-quasi-LL.
In the limit of $R \ll l$, Eq.~(\ref{eq:CFLLtheta}) reduces to
\be
\tilde{\theta}^* = -1 + \frac14 \left(\frac R l\right)^2 
+ {\cal O}\left(\frac R l\right)^4,
\ee
showing a quadratic increase from $-1$.  
Similar behavior is displayed by two electrons in the 
second Landau level for small
separations.  For $\nu=1$, the statistics parameter for electrons 
separated by a distance $d$ is given by
\be
\tilde{\theta}^* = - \frac{(d/l)^2}{2(1- e^{-d^2/2l^2})}.
\ee
This yields in the limit $d \ll l$ 
\be
\tilde{\theta}^* = -1 + \frac14 \left(\frac d l\right)^2 
+ {\cal O}\left(\frac d l\right)^4 ,
\ee
which is identical to that for composite fermions in lowest CF-quasi-LL.

We can expect similar behavior for CFQPs in higher CF-quasi-LLs.
The only difference is that they feel an effective magnetic field
$B^* = B/(2pn+1)$, which changes the length scale from $l$ to $l^*$.
It is expected that for small $d$, $\tilde{\theta}^*$ is given by
\be
\tilde{\theta}^* = -1 + \frac1{4(2pn+1)} \left(\frac{R}{l}\right)^2 
+ {\cal O}\left(\frac R l\right)^4 \;.
\label{eq:near}
\ee
Figure~\ref{fig:near} presents the calculated $\tilde{\theta}^*$ for small
separation at $\nu=1/3$ and 2/5 along with the heuristic expression of  
Eq.~(\ref{eq:near}).
As can be seen in Fig.~\ref{fig:near}, Eq.~(\ref{eq:near}) 
gives a good account of the behavior at both fillings. 

\subsection{CF quasiparticles in different CF quasi-Landau levels}

In this section we investigate another interesting question: What is the 
relative statistics for two CFQPs in different 
CF-quasi-Landau levels?  This corresponds to the situation when 
a CFQP is inserted into an excited CF-quasi-Landau level.  From the CF 
point of view, the statistics is related to the excess charge
due to the presence of the additional CFQP as shown in
Eq.~(\ref{eq:DeltaPhi}).
Since the local charge of the CFQP is independent of the quasi-Landau level to
which it belongs, the resulting statistics is expected to the same as
that when both CFQPs are in the same CF quasi-Landau levels.

For an explicit calculation,
we investigate the situation that a CFQP in the second
CF-quasi Landau level goes around a CFQP in the third CF-quasi Landau level 
at the filling for $\nu=1/3$.
The wave function for two CFQPs is given by
\bea
\Psi^{\eta,\eta'}_{1/3} &=& {\cal P} 
\left|\begin{array}{ccccc}
\phi^{(1)}_{\eta}(\vec{r}_1) & \phi^{(1)}_{\eta}(\vec{r}_2) & . &.&.\\
\phi^{(2)}_{\eta'}(\vec{r}_1) & \phi^{(2)}_{\eta'}(\vec{r}_2) & . &.&.\\
1 & 1 & . &.&. \\
z_{1}&z_{2}&. &.&.\\
.&.&.&.&.\\
.&.&.&.&. \\
z_1^{N-3} & z_2^{N-3} & . &.&. 
\end{array}
\right| \nonumber \\
& & \;\;\;\;\times \prod_{i<k=1}^N(z_i-z_k)^{2} e^{-\sum_j|z_j|^2/4l^2}\;.
\label{2CFQPdiff}
\end{eqnarray}
For simplicity, we set $\eta'=0$.

\begin{figure}
\centerline{\epsfig{file=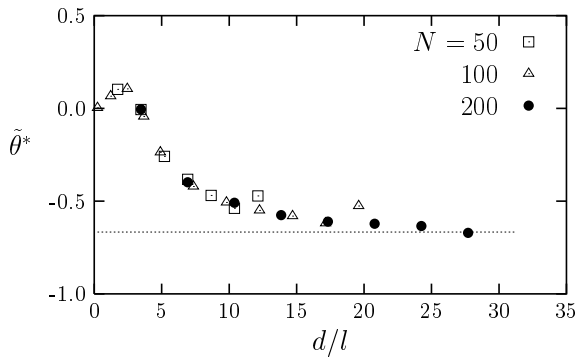,width=3.0in,angle=0}}
\caption{The statistical angle $\tilde\theta^*$ for the CFQPs 
in different CF-quasi-LLs at the filling $\nu=1/3$ 
as a function of $d \equiv |\eta-\eta'|$.  
The CFQP at the origin is in the third CF-quasi-LL while the CFQP traversing 
a closed loop is in the second CF-quasi-LL.
$N$ is the total number of composite fermions, and $l$ is the magnetic length. 
}
\label{fig:DiffLevel}
\end{figure}

Figure~\ref{fig:DiffLevel} demonstrates that the asymptotic value of the 
relative statistics of two CFQPs in two different CF-quasi-Landau levels is 
the same as for those in the same CF-quasi-Landau level.  On the other hand,
there is significant difference for small separations between two
CFQPs. The behavior at small separations is believed to be 
sensitive  to the local structure of each CFQP, because corrections to the 
statistics are caused by their overlap.

Jeon and Jain~\cite{Quasiparticles} noted that for two 
quasiparticles at the origin
there is a qualitative difference between the wave functions constructed  
according to Laughlin's ansatz and the one used above based on the CF theory at 
$\nu=1/3$.
For CFQPs there are many candidates for two quasiparticle states.
The CF $[N-2,2]$ with both two quasiparticles in the second CF-quasi-LL 
has lowest energy among the candidates as expected from
the fact that it has the lowest effective cyclotron energy. 
As discussed in Ref.~\onlinecite{Quasiparticles}, 
Laughlin's wave function for two 
quasiparticles is more akin to the $[N-2,1,1]$ state of composite fermions,
with one CFQP in the second CF-quasi Landau level and the other in the third;  
both states has the same total angular momentum and their density
profiles look alike.  One might therefore have expected that 
the $[N-2,1,1]$ state would not display definite statistics.
However, our result above demonstrates that even the $[N-2,1,1]$ state 
is fundamentally different from the one in Eq.~(\ref{L2q}).

\subsection{Composite fermions: fermions or anyons?}

The fractional statistics of the CFQPs ought not to be confused with 
the fermionic statistics of composite fermions.  The wave 
functions of composite fermions 
are single-valued and antisymmetric under particle exchange;
the fermionic statistics of composite fermions has been 
firmly established through a variety of facts, including the observation 
of the Fermi sea of composite fermions, the observation of FQHE at fillings 
that correspond to the IQHE of composite fermions, and also by the fact that 
the low energy spectra in exact calculations on finite systems 
have a one-to-one correspondence with those of weakly interacting
fermions.~\cite{Heinonen}  The appearance of fractional statistics may seem at 
odds with the fermionic nature of composite fermions, but there is 
no contradiction.  After all, any fractional statistics in 
nature {\em must} arise in a theory of particles that are either fermions or bosons
when an ``effective" description is sought in terms of a small number of 
collective degrees of freedom.  The fractional statistics appears 
in the CF theory when all of the original particles at $\{z\}$ are integrated out 
(or treated in an average, mean field sense) to formulate an effective 
description in terms of the few CFQPs at $\{\eta\}$.  If we 
work with {\em all} composite fermions, then Eq.~(\ref{phi*}) is sufficient.

\subsection{Constraints on possible observation of fractional statistics}

There are features that complicate a possible observation  
of fractional statistics.  (i)  The CFQPs 
are not {\em ideal} anyons.  As seen in our calculations, the 
fractional statistics is sharply defined
only asymptotically; in general there are corrections to it.
Substantial deviation of $\theta^*$ from its asymptotic
value is seen at separations of up to $10$ magnetic lengths.
Therefore, a measurement of $\theta^*$ must ensure that 
there is no overlap between the CFQPs at any time.
One might expect that the interaction between the 
CFQPs will be repulsive which will automatically ensure that 
they do not come very close to one another.  That turns out 
not to be the case, however.  The interaction between the CFQPs 
is very weak and often {\em attractive}.~\cite{Lee}
(ii)  There is another important aspect through which the 
situation here differs from that for ideal anyons.  
For two ideal anyons, the Berry phase is zero for paths with 
zero winding number and $2\pi \theta^*$ for paths with 
unit winding.  One therefore only needs to measure the 
Berry phase for a path that encircles another particle.
In the case of the FQHE, on the other hand, the 
fractional statistics, itself an 
O(1) quantity, arises as a difference between two O($N$)  
Berry phases, where $N$ is the number of particles enclosed by 
the closed trajectory.  For the reason listed in (i), $N$ 
must necessarily be quite large.  A precise measurement of the difference 
therefore requires an almost perfect control over the trajectory.
Fluctuations in the trajectory
on the order of the size of the CFQP will produce 
O($\sqrt{N}$) fluctuations in each Berry phase which will  
completely obscure the O(1) difference.
(Our calculation actually provides an example where an immeasurable 
error in the trajectory produces a finite correction to 
$\theta^*$, changing its sign.) In fact, one may ask how quantum fluctuations in each 
O($N$) quantity affect the O(1) difference, and whether 
the O(1) difference can be defined in a rigorous manner.~\cite{Comment2} 
(In this context, it is noted that the effective magnetic field
is related to the total Berry phase associated with
a path, an order $N$ quantity, and therefore robust to
quantum mechanical fluctuations which are of smaller order.)
(iii) There are many other features likely to be 
present in a real experimental situation that 
would be inimical to an observation of fractional 
statistics, for example, disorder and finite temperature, both of 
which generate particle-hole pairs which would provide 
a correction.  (iv) The current flows at the edge of an incompressible 
FQHE system, where the fractional statistics is not well defined due 
to the absence of a gap.~\cite{GJJ} This creates a problem for a 
detection of fractional statistics in a transport experiment. 
(v) It is not known how robust the fractional statistics concept 
is to perturbations.  We have confirmed it for {\em non-interacting}
composite fermions.  However, it has been found that interactions 
between composite fermions can produce significant corrections to 
apparently topological quantities.~\cite{Mandal}
Also, the fact that certain quasiparticle  
wave functions at $\nu=1/m$ do not produce a sharp fractional statistics
shows that it is not as robust as the fractional charge or 
the effective magnetic field.  Whether it survives a more realistic 
calculation remains to be tested.

\begin{acknowledgments}
Partial support of this research by the National Science Foundation under grants
no. DGE-9987589 (IGERT) and DMR-0240458 is gratefully acknowledged.
We thank Profs. A.S. Goldhaber and J.M. Leinaas for comments.
\end{acknowledgments}


\begin{thebibliography}{99}
\bibitem{Comment}  It was shown that the 
fractional statistics can be ``derived" heuristically if 
one {\em assumed} incompressibility at a fractional filling of the
lowest LL, and made certain other natural assumptions.~\cite{Su}
However, the fact that Laughlin's model for the quasiparticles does not 
produce the expected fractional statistics provides a contradiction.  
It is not known which of the assumptions in the derivation is invalid 
for that case.
\bibitem{Comment2}
Such subtleties appear also in the definition of fractional charge.
There is no fundamental particle with a fractional charge; the fractional
charge is the difference between the net charge in a large enough 
area with or without a CFQP in it.  In other words, it is 
also an order one difference between two large numbers.  Will quantum 
fluctuations in the net charge, which are large compared to unity for a large 
area, make the definition of a fractional charge conceptually meaningless?
Here, a possible way out~\cite{charge} is to consider an area enclosing the 
CFQP which is small enough that the fluctuations are 
small compared to the difference.  Any experiment seeking 
to measure the fractional charge must consider such issues carefully.
\end{thebibliography}
\end{document}